\documentclass{article}

\usepackage{arxiv}

\usepackage[utf8]{inputenc} 
\usepackage[T1]{fontenc}    
\usepackage{hyperref}       
\usepackage{url}            
\usepackage{booktabs}       
\usepackage{amsfonts}       
\usepackage{nicefrac}       
\usepackage{microtype}      
\usepackage{graphicx}
\usepackage{amssymb}
\usepackage{amsmath}
\usepackage{lineno}
\usepackage{hyperref}
\usepackage{listings}
\usepackage{subfigure}
\usepackage{comment}
\usepackage{enumitem}
\usepackage{multirow}
\usepackage[dvipsnames]{xcolor}

\newcommand\norm[1]{\left\lVert#1\right\rVert}

\DeclareMathOperator*{\argmin}{arg\,min}

\title{Seeing through the CO$_2$ plume: joint inversion-segmentation of the Sleipner 4D Seismic Dataset}

\author{
  Juan Romero \\
  KAUST\\
  Thuwal, Kingdom of Saudi Arabia \\
  \texttt{juan.romeromurcia@kaust.edu.sa}\\
   \And
  Nick Luiken \\
  KAUST\\
  Thuwal, Kingdom of Saudi Arabia \\
  \texttt{nicolaas.luiken@kaust.edu.sa}\\
  \And
  Matteo Ravasi \\
  KAUST\\
  Thuwal, Kingdom of Saudi Arabia \\
  \texttt{matteo.ravasi@kaust.edu.sa}\\
  }
  
\begin{document}

\chead{Seeing through the CO$_2$ plume: joint inversion-segmentation of the Sleipner 4D Seismic Dataset}

\maketitle

\begin{abstract}
    4D seismic inversion is the leading method to quantitatively monitor fluid flow dynamics in the subsurface, with applications ranging from enhanced oil recovery to subsurface CO$_2$ storage. The process of inverting seismic data for reservoir properties is, however, a notoriously ill-posed inverse problem due to the band-limited and noisy nature of seismic data. This comes with additional challenges for 4D applications, given inaccuracies in the repeatability of the time-lapse acquisition surveys. Consequently, adding prior information to the inversion process in the form of properly crafted regularization terms is essential to obtain geologically meaningful subsurface models. Motivated by recent advances in the field of convex optimization, we propose a joint inversion-segmentation algorithm for 4D seismic inversion, which integrates Total-Variation and segmentation priors as a way to counteract the missing frequencies and noise present in 4D seismic data. The proposed inversion framework is applied to a pair of surveys from the open Sleipner 4D Seismic Dataset. Our method presents three main advantages over state-of-the-art least-squares inversion methods: 1. it produces high-resolution baseline and monitor acoustic models, 2. by leveraging similarities between multiple data, it mitigates the non-repeatable noise and better highlights the real time-lapse changes, and 3. it provides a volumetric classification of the acoustic impedance 4D difference model (time-lapse changes) based on user-defined classes, i.e., percentages of speed-up or slow-down in the subsurface. Such advantages may enable more robust stratigraphic/structural and quantitative 4D seismic interpretation and provide more accurate inputs for dynamic reservoir simulations. Alongside presenting our novel inversion method, in this work, we introduce a streamlined data pre-processing sequence for the 4D Sleipner post-stack seismic dataset, which includes time-shift estimation and well-to-seismic tie. Finally, we provide insights into the open-source framework for large-scale optimization that we used to implement the proposed algorithm in an efficient and scalable manner.
\end{abstract}

\section{Introduction}

The Sleipner CO$_2$ project in the Norwegian North Sea is the world's first commercial carbon storage project aimed at carbon emission mitigation. More than 19 million by-product tonnes of CO$_2$ from gas production in the Sleipner West field have been injected into the shallower Utsira Formation since 1996. To understand the behavior of the injected CO$_2$ in the subsurface and to prevent potential geohazards, a continuous seismic monitoring program with nine seismic surveys has taken place around the injection area \cite{furre2017}. The injected CO$_2$ in the subsurface is observed in post-stack seismic data as a series of high amplitude reflections that expand laterally over time \cite{chadwick2010}. Such anomalies present amplitude variations and suffer from tuning effects due to the thin layering of sands and shales in the Utsira Formation \cite{Furre2015}. These characteristics motivate the application of 4D seismic inversion techniques to better interpret the lithological and fluid-related seismic effects. Pioneering work in 4D seismic inversion of the Sleipner data has been carried out on post-stack data \cite{Ghosh2015}, pre-stack data \cite{Ghaderi2009}, and waveform data \cite{Dupuy2017}, showing that it is possible to estimate reliable reservoir property changes through time. However, the presence of non-repeatable noise in the recorded seismic data affects the performances of these methodologies, as the noise is usually carried over to the inversion products, possibly hindering the interpretation of real geological time-lapse changes.\\

In this paper, we propose to apply a 4D joint inversion-segmentation (JIS) algorithm, originally introduced by \cite{Romero2022}, to two vintages of the 4D Sleipner seismic dataset. The initial model used in JIS is obtained from the time-converted and calibrated impedance well-log, which is also used as input to a seismic-to-well tie procedure. Finally, the baseline and monitor seismic data are aligned prior to inversion using time-shifts estimated by means of non-linear inversion using the Gauss-Newton scheme. Our method is compared to state-of-the-art least-squares inversion using generalized Tikhonov regularization that is used to enforce smoothness for both the baseline and monitor acoustic impedance models as well as their difference. Moreover, we discuss the added value introduced by the segmentation step in our 4D JIS algorithm, which results in a volumetric representation of the key expected time-lapse changes in the subsurface. Finally, owing to the increased computational cost of the 4D JIS algorithm compared to the state-of-the-art least-squares method and given the large dimensions of the Sleipner seismic dataset, we provide insights into our open-source, GPU-based implementation of the inversion and segmentation algorithms. A link to the accompanying codes can be found in the Complementary material section.

\subsection{4D Post-stack seismic inversion}
Post-stack seismic inversion aims to retrieve the natural logarithm of the acoustic impedance model $m$ from post-stack seismic data $d$ by solving the following optimization problem:

\begin{equation}
   \argmin_m \frac{1}{2}\Vert Gm - d\Vert_2^2 \quad \quad \text{where} \quad G = W D 
\label{eqn:ip}
\end{equation}

where the post-stack modeling operator $G$ is defined as the chain of a time derivative operator $D$ and a convolution operator $W$ created for a given wavelet $w$ \cite{stolt1985}. In the case of 4D post-stack seismic inversion, baseline $m_1$ and monitor $m_2$ acoustic impedance models are estimated simultaneously through the joint inversion of the baseline $d_1$ and monitor $d_2$ seismic data:

\begin{eqnarray}
    \argmin_{m_1, m_2} \frac{1}{2} \left\Vert\left[\begin{array}{ll}
    G_1 & 0 \\
    0 & G_2 \\
    \end{array}\right]
    \left[\begin{array}{ll}
    m_1  \\
    m_2  \\
    \end{array}\right]
     - \left[\begin{array}{ll}
    d_1  \\
    d_2  \\
    \end{array}\right]
    \right\Vert_2^2
    \hspace{0.4cm} := \hspace{0.4cm} 
    \argmin_{\widetilde{m}} \frac{1}{2}\Vert \widetilde{G}\widetilde{m} - \widetilde{d}\Vert_2^2  
\label{eqn:4d}
\end{eqnarray}

where $G_1$ and $G_2$ are the baseline and monitor modeling operators, respectively. Note that the two operators might have different wavelets $w_1$ and $w_2$.

\subsubsection{Regularization}
Given the limited frequency range associated with a seismic wavelet, the operator $G$ acts as a band-limiting operator: this behaviour makes seismic inversion an ill-posed inverse problem. Additionally, the acquired data contain noise that is non-repeatable across surveys: if not handled properly, such noise inevitably carries over to the inverted models. Therefore, the data misfit term in equation \ref{eqn:4d} is routinely augmented with one or more regularization terms that, on the one hand, encode our prior knowledge of the structure of the subsurface and, on the other hand, allow to couple the models using a-priori information of the time-lapse effects that we wish to recover. More specifically, the coupling term strives to filter out non-repeatable noise whilst enhancing changes in acoustic impedance that make geological sense. A widely used strategy in industry involves introducing Tikhonov regularization for both the independent models and their 4D difference:

\begin{equation}
  \argmin_{\widetilde{m}}\quad \frac{1}{2}\norm{ \widetilde{G} 
    \widetilde{m}
    - 
    \widetilde{d}}_2^2 
    + 
    \alpha 
    \norm{ 
    \widetilde{\nabla}^2 
    \widetilde{m}}_2^2
    + 
    \beta
    \norm{\Delta \widetilde{m}}_2^2.
\label{eqn:ip2}
\end{equation}

Here, $\widetilde{\nabla}^2=\text{diag}\{\nabla^2, \nabla^2\}$, $\nabla^2$ is a two-dimensional Laplacian operator, $\Delta \widetilde{m} = m_2 - m_1$ (or vice versa, based on the adopted convention for 4D changes), and $\alpha$ and $\beta$ are regularization parameters. The first regularization term encourages smooth impedance models, whilst the second regularization term enforces the difference between baseline and monitor to be small. Due to the layered structure of the subsurface a more appropriate prior for the impedance models is represented by Total-Variation (TV) regularization, which promotes sharp jumps across interfaces thereby retrieving higher resolution subsurface models \cite{Romero2022}.

\subsubsection{4D Joint inversion and segmentation (JIS)}
In this paper, we propose the use of the JIS algorithm \cite{ravasi2022, Romero2022, romero2023}, which in addition to TV regularization terms, it incorporates a segmentation constraint over the estimated 4D difference. As such, 4D JIS not only retrieves high-resolution baseline and monitor acoustic impedance models, but also segments the monitor-baseline acoustic impedance difference into multiple probability volumes based on user-defined time-lapse classes. This is  achieved in practice by solving the following bi-objective optimization problem:
\begin{eqnarray}
    \argmin_{\widetilde{m}, V} \frac{1}{2} \left\Vert
    \widetilde{G} \widetilde{m} - \widetilde{d}
    \right\Vert_2^2 +
    \alpha \Vert \widetilde{m} \Vert_{TV} + 
    \delta \sum_{j=1}^{N_c} \sum_{i=1}^{N_x N_y N_z}  V_{ji} (\Delta m - c_j)^2 +\beta \sum_{j=1}^{N_c}\Vert V^T_j \Vert_{TV}.
\label{eqn:jis}
\end{eqnarray}
Here, $(N_x, N_y, N_z)$ are the 3D model dimensions, $V$ is the segmentation matrix whose $V_j$ rows contain the probability of each model's cell to belong to a certain class $c_j$, and $\Vert\cdot\Vert_{TV}$ denotes the isotropic Total-Variation norm. Finally, $\alpha, \delta$, and $\beta$ are the weights associated with the regularization and segmentation terms. In this paper, we define three classes, namely $c_1 = -50\%$ indicating a decrease in acoustic impedance (or slow-down), $c_2= 0\%$ referring to no impedance changes, and $c_3 = 50\%$ corresponding to  an increase in acoustic impedance (or speed-up). The third term in equation \ref{eqn:jis} corresponds to the segmentation term and aids the inversion process by providing information about the expected 4D changes, whilst the fourth term ensures that the segmented partitions have smooth perimeters. Equation \ref{eqn:jis} presents a non-convex optimization problem that becomes convex when fixing one of the optimization variables $(\widetilde{m}, V)$. Therefore its solution is found in an alternating fashion, where each variable is optimized independently at every iteration using the Primal-Dual algorithm \cite{Chambolle2011}; for more implementation details, we refer the reader to \cite{ravasi2022}.

\section{Methodology}
\subsection{The Sleipner dataset}
\begin{figure*}[!h]
    \centering
    \includegraphics[width=\textwidth]{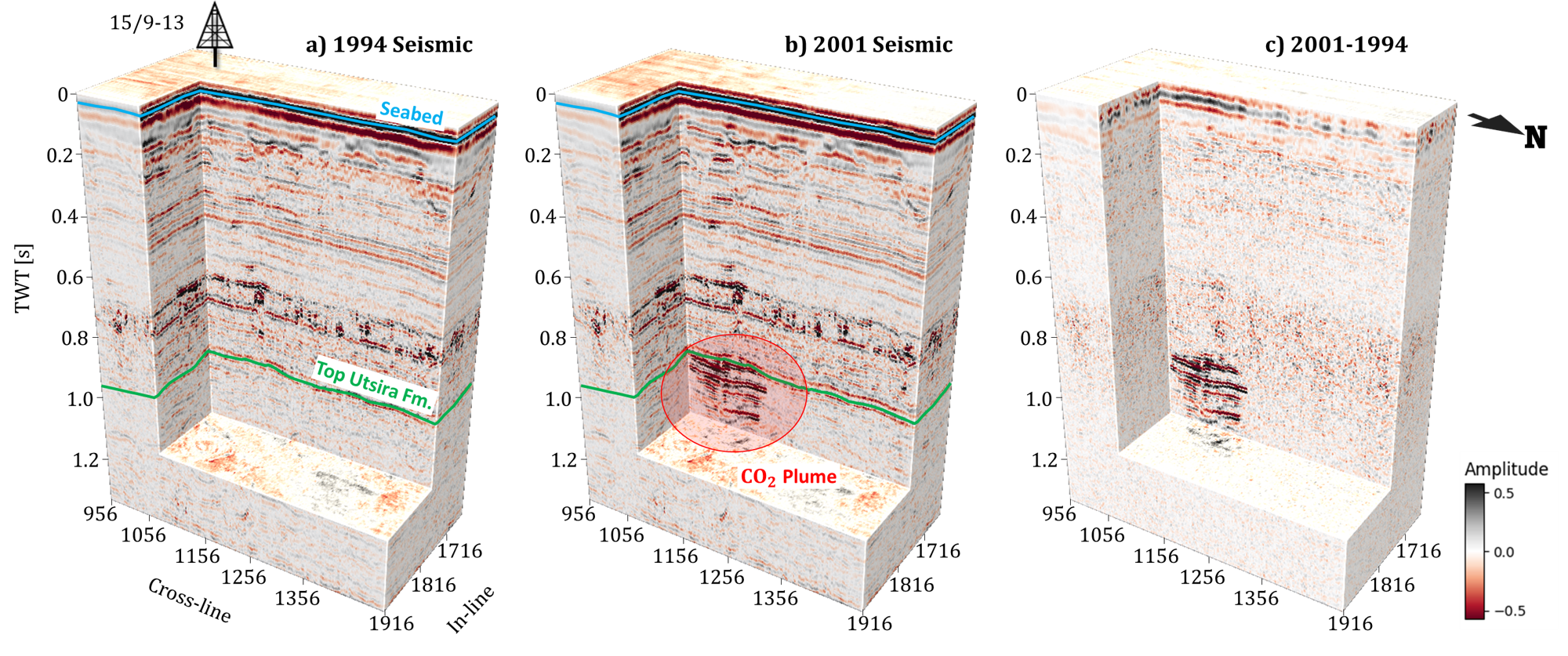}
    \caption{Sleipner seismic dataset used in this work. a) Baseline seismic survey (1994), b) monitor seismic survey (2001), and c) the difference between them. The well shown in a) does not intersect the CO$_2$ plume.}
    \label{fig:sleipnerdata}
\end{figure*}

The publicly available Sleipner dataset contains a 4D Seismic dataset that is composed of one baseline seismic survey acquired in 1994 and six monitor seismic surveys acquired in 1999, 2001, 2004, 2006, 2008, and 2010. This extensive seismic acquisition programme has been established to continuously monitor the injection of CO$_2$ into the saline aquifer of the Middle Miocene/Early Pliocene Utsira Formation \cite{chadwick2010}. The seismic data cover an area of approximately 4$\times$7 km$^2$, with a record length of 2 seconds and a sample rate of 2 ms. For this work, we use the 1994 and 2001 vintages from the latest re-processing in 2010 (Figure \ref{fig:sleipnerdata}) and consider a subvolume with same spatial dimensions as the original and a duration of 1.36 seconds (part of the underburden is removed). The open Sleipner dataset also contains data from 2 wells: 15/9-A-16 and 15/9-13. Well 15/9-13 is a nearly vertical well that does not intersect the CO$_2$ plume and contains a complete set of well-logs that includes a check-shot calibrated sonic well-log. Well 15/9-A-16 is a horizontal injection well lacking sonic log and check-shot, and it is therefore not used in this study.

\begin{figure*}[!h]
    \centering
    \includegraphics[width=\textwidth]{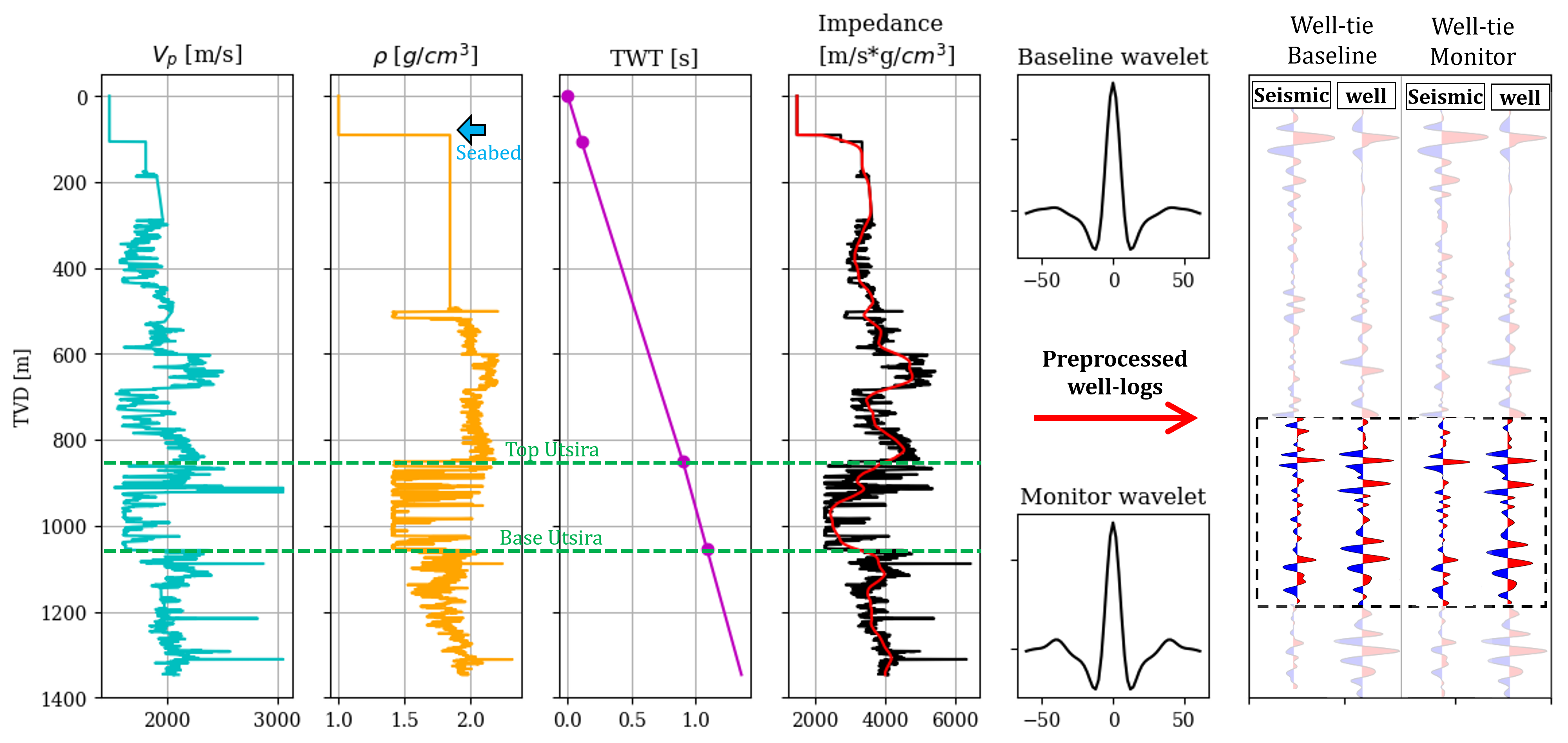}
    \caption{Well-analysis and well-to-seismic tie. From left to right, sonic and density logs (edited to eliminate spikes), TD-TWT relationship obtained by correlating regional seismic surfaces in time with well-picks in depth, impedance log (black) and its  smoothed version used to create the background model for inversion (red), estimated wavelets, and real and synthetic seismic traces at well location for both the baseline and monitor seismic surveys, the latter computed with amplitude-calibrated statistical wavelets. Taken from \cite{romero2023}.}
    \label{fig:welllogs}
\end{figure*}

To begin with, we conduct the well-to-seismic tie using as input an edited version of the check-shot corrected sonic log and the density log for Well 15/9-13 (Figure \ref{fig:welllogs}). We create the time-depth relationship by linking well-picks and regional time reflectors identified in the baseline seismic volume for the seabed and the top and base of the Utsira Formation. Then, we estimate two wavelets, one per survey, using a statistical approach based on the seismic amplitude spectrum of an extracted sub-volume around the reservoir depth. Each wavelet is scaled accordingly by comparing the modelled synthetic trace at the well location with the closest trace in the seismic data. When convolving the wavelet with the derivative of the impedance well-log, we obtain the modeled baseline and monitor synthetic traces (figure \ref{fig:welllogs}). We observe a fair correlation at the reservoir depth when comparing the modeled traces to the closest to well seismic baseline and monitor traces. The resulting time-converted impedance log is smoothed to be used as the initial inversion model.

\subsection{Time-shift inversion}

In the context of time-lapse seismic, the term time-shift is commonly used when referring to the amount of time mismatch between equivalent reflectors (hence geological events) in the baseline and the monitor datasets. Such time-shifts are caused by velocity changes in the subsurface due to geomechanical effects (e.g., strain changes in the overburden and/or reservoir) and/or dynamic fluid saturation \cite{Macbeth2019}, which are unaccounted for in the migration velocity model of the monitor datasets. The relationship between equivalent traces in the baseline $b(t)$ and monitor $m(t)$ seismic dataset can be described as
\begin{equation*}
    b(t) \approx m(t + \tau (t)),
\end{equation*}
where $\tau(t)$ is the time-shift field. In this work, we use the nonlinear inversion approach proposed by \cite{Ricket2007} to find a time-shift volume across the entire area of interest. This is achieved by solving a sequence of linearized inverse problems via Gauss-Newton iterations as follows:
\begin{equation*}
     \Delta \tau_i = \argmin_{\Delta \tau} ||b(t) - (m(t+ \tau_{i-1}) + J_m\Delta \tau)||^2 + \epsilon^2 ||\nabla^2 (\tau_{i-1} + \Delta \tau)||^2, \quad  \tau_i =  \tau_{i-1} + \Delta \tau_i.
\end{equation*}
Here, $J_m = -\text{diag}\{\frac{db}{dt}|_{t=t+\tau_{i-1}}\}$ is the Jacobian matrix, $\nabla^2$ is the Laplacian operator and $\epsilon$ is the regularization parameter. The regularization term ensures smooth time-shift estimates in both the spatial and time axes. This non-linear inversion scheme is the most prominent method for time-shift estimation \cite{Macbeth2019}: its main advantage against methods based on cross-correlation is that it can be applied globally, and the estimated time-shift field can be smoothed (regularized) over all the dimensions in the volume of interest. A limitation of this approach is that it works well only  for small time-shifts (i.e., less than half of the dominant period of the seismic data) due to the linearization introduced to solve the problem. Otherwise, it might suffer from linearization errors. In the case of Sleipner seismic datasets, the dominant period is around 0.04 ms, and therefore reliable time-shift values are only those under 20 ms.

\begin{figure*}[!h]
    \includegraphics[width=15.6cm]{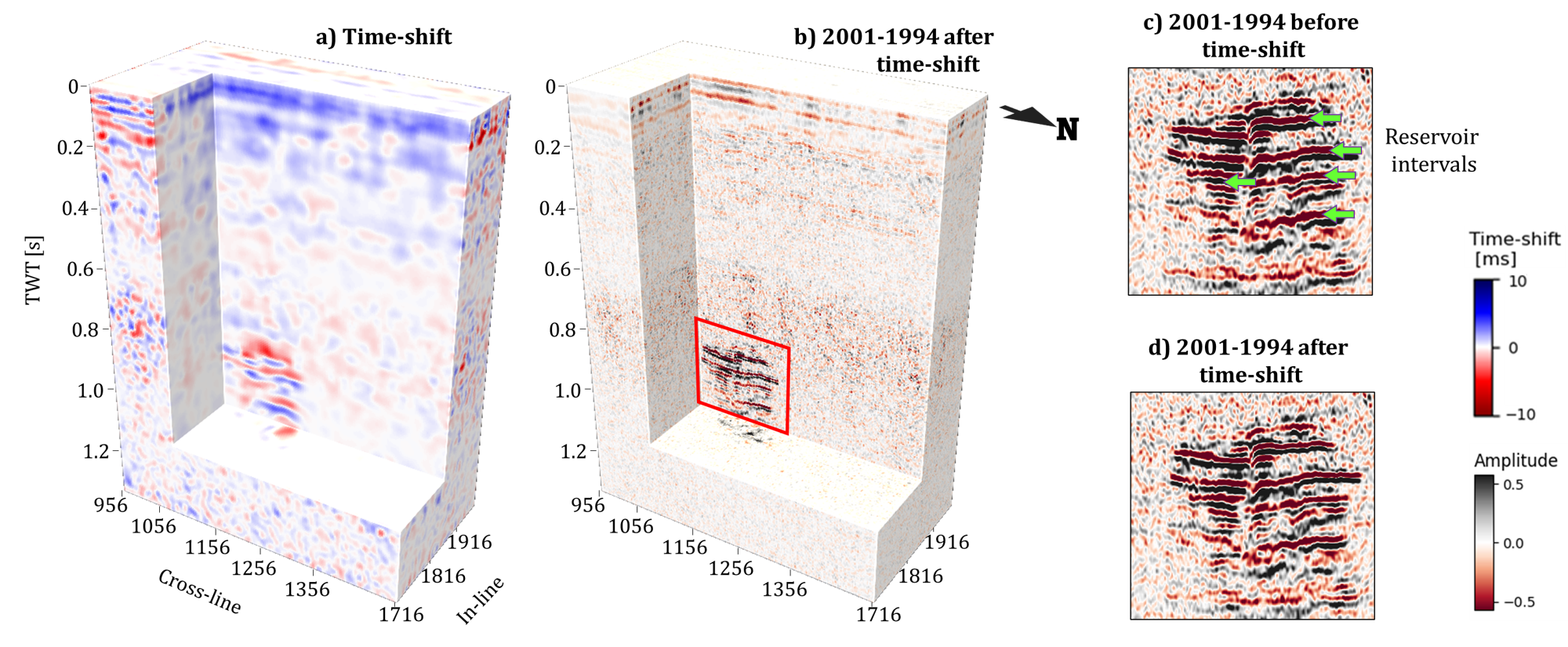}
    \caption{Time-shift processing step. a) Estimated time-shift volume, b) 2001-1994 difference after correcting the 2001 dataset for the estimated time-shifts. c) and d) Close-ups from the red square rectangle at b) of the difference volumes at the reservoir level before and after time-shift correction. Green arrows highlight the reservoir reflections.}
    \label{fig:timeshift}
\end{figure*}

The estimated time-shift volume (Figure \ref{fig:timeshift}a) shows a layered structure of speed-ups combined with slow-downs at the reservoir level with peak values around $10ms$; such values are in agreement with previously published time-shift estimates for this dataset \cite{Furre2015}, which also observed that some of the time-shifts at the reservoir need further calibration to account for wavelet distortions. The overburden shows low and unstructured time-shifts, likely caused by non-repeatable noise in the matched datasets. On the other hand, a nearly homogeneous trend of positive time-shifts is observed in the underburden. This time-shift distribution between the overburden, reservoir, and underburden may be interpreted as velocity changes that occurred only at the reservoir interval as a consequence of the injection of the less dense CO$_2$ into previously brine-filled layers \cite{Macbeth2019}. The layered structure in the reservoir interval might in fact suggest the presence of geomechanical deformation in the interbedded shale layers; this is however likely not the case and can be misleading as these values need to be corrected for tuning effects.

Once the monitor seismic dataset is corrected for the estimated time-shifts (Figure \ref{fig:timeshift}b), differences in the overburden, like near the seabed, are significantly reduced, providing a cleaner 4D difference volume (compare Figures \ref{fig:sleipnerdata}c and \ref{fig:timeshift}b), which is of great value for any subsequent qualitative or quantitative interpretation step. When looking at the reservoir level (Figures \ref{fig:timeshift}c and \ref{fig:timeshift}d), we observe how, after time-shift correction, some of the small-scale reservoir reflections (green arrows on figure \ref{fig:timeshift} c) observed as troughs in the original 4D difference data become thinner and better defined in the corrected one.

\subsection{Efficient large-scale optimization}

\subsubsection{Linear Operators and matrix-free based inversion} Solving linear inverse problems entails creating a dense matrix that represents the physical interaction between the model and data vectors. In practical large-scale applications, this approach quickly becomes unfeasible when dealing with millions of model and data parameters. Luckily, one can usually identify sparsity or structure in most scientific problems and rely on so-called \textit{linear operators}: from a computational point of view, a linear operator can be identified as a pair of routines that compute the action of the modelling operator (and its adjoint) on a vector, rather than computing the matrix-vector product explicitly. As an example, let's take the first-order finite-difference operator, where the associated matrix contains  non-zero elements along two diagonals. In this case, the action of the operator can be far more efficient if implemented by directly computing the weighted difference between subsequent elements of the input vector. PyLops\footnote{\url{http://github.com/pylops/pylops}} \cite{Ravasi2020} is a Python library designed around this concept, aimed at solving large-scale optimization problems in an efficient manner. More specifically, the key idea behind PyLops is to represent any linear operator as a Python class-based entity whose key methods perform the forward (e.g., matrix-vector or matrix-matrix products) and adjoint operations. This happens to be beneficial as any iterative solver that is used to solve large systems of equations relies on matrix-vector products rather than on the matrix itself. Apart from being usually more efficient than their dense counterparts, another key advantage of class-based linear operators is the reduction in memory footprint. This ultimately allows, in many cases, to apply these operators on accelerators (i.e., GPUs), as discussed in more detail in a later section. Moreover, PyLops operators are designed in a modular fashion such that different operators can be arranged and combined seamlessly. For example, in the seismic inversion problem, we can easily combine some of PyLops' core operators as \textit{`lego blocks'} to create the post-stack modelling operator as well as the various regularization terms. This represents a major difference when compared to other open-source frameworks available to the geophysical community (e.g., Seismic Unix, Madagascar), which take instead an application-centric approach and are far less flexible when it comes to building new applications for research and/or production purposes.

\subsubsection{Proximal Operators} Proximal algorithms \cite{parikh2014} are a class of optimization algorithms that are generally used to minimize non-smooth functionals, an example being TV-regularized seismic inversion. Examples of popular proximal algorithms are the Fast Iterative Shrinkage-Thresholding Algorithm (FISTA -- \cite{Beck2009}), the Alternating Direction Method of Multipliers (ADMM --  \cite{Boyd2011}), and the Primal-Dual algorithm. Similar to gradient-based algorithms, these solvers are iterative in nature and typically involve a gradient-step followed by a so-called proximal operator step (which we can loosely speaking identify as the counterpart of the gradient for non-smooth functionals). In this work, we leverage PyProximal\footnote{\url{http://github.com/pylops/pyproximal}}, a python library under early development that provides an extensive set of proximal operators and solvers and allows for a direct integration with PyLops' linear operators. More specifically, for our problem of interest, we use the Primal-Dual algorithm \cite{Chambolle2011} to solve the TV-regularized 4D seismic inversion in equation \ref{eqn:4d} since this has been shown to outperform other proximal algorithms in seismic inversion \cite{ravasi2022, Romero2022}.

\subsubsection{GPU acceleration} 
\begin{figure*}[!h]
    \centering
    \includegraphics[width=7cm]{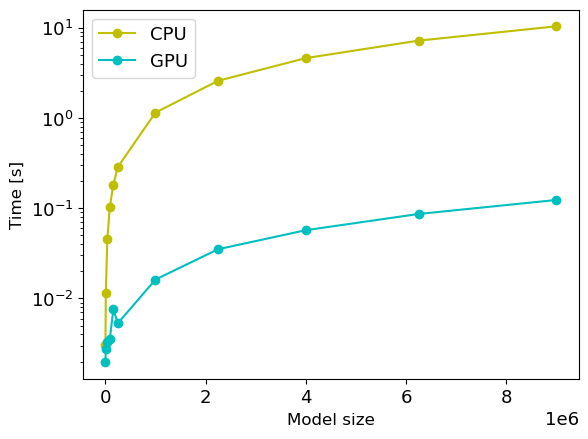}
    \caption{Computational time of a GPU compared with a CPU to solve a bisection problem for different model sizes.}
    \label{fig:gpu}
\end{figure*}

In recent years, machine learning has become increasingly popular, with several successful applications in science and engineering. GPUs have played a pivotal role, especially in the surge of deep learning, because of their superiority over CPUs in training large neural networks. Such a superiority stems from the fact that GPUs are much faster than CPUs at carrying out matrix-matrix and matrix-vector multiplications (and other key scientific computing kernels -- e.g., convolutions); this feature makes GPUs also very suitable for large-scale inverse problems. CUDA is the most popular programming language for GPUs: whilst originally designed to work with programming languages such as C, C++, and Fortran, more recently, this programming model has also been exposed to Python through just-in-time compiler libraries such as Numba. Moreover, most of the core linear algebra and signal processing routines provided by CUDA libraries, such as cuBLAS and cuFFT, can be easily accessed nowadays within Python through the CuPy library. In this work, we leverage the power of GPUs at two stages of the inverse process: first, both the post-stack modelling operator and the gradient operator involved in the TV regularization term operate directly on GPU arrays, making the seismic inversion process orders of magnitude faster than its CPU-based version. Second, because of its embarrassingly parallel nature, the segmentation task is also performed on GPUs using a custom CUDA kernel written in Numba: this is mainly due to the fact that the proximal operator involved in the segmentation process requires the solution of a bisection problem for every grid point independently. In Figure \ref{fig:gpu}, we compare CPU and GPU computation time for a synthetic segmentation problem for different model sizes. It can be observed that the bisection algorithm's computation for the segmentation of a million-size seismic dataset takes around 10 seconds on the CPU and 0.1 seconds on GPU. All experiments are run on an AMD EPYC 7713P 64-Core Processor equipped with a single NVIDIA TESLA A100. 

\section{Results}
\begin{figure*}[!h]
    \centering
    \includegraphics[width=\textwidth]{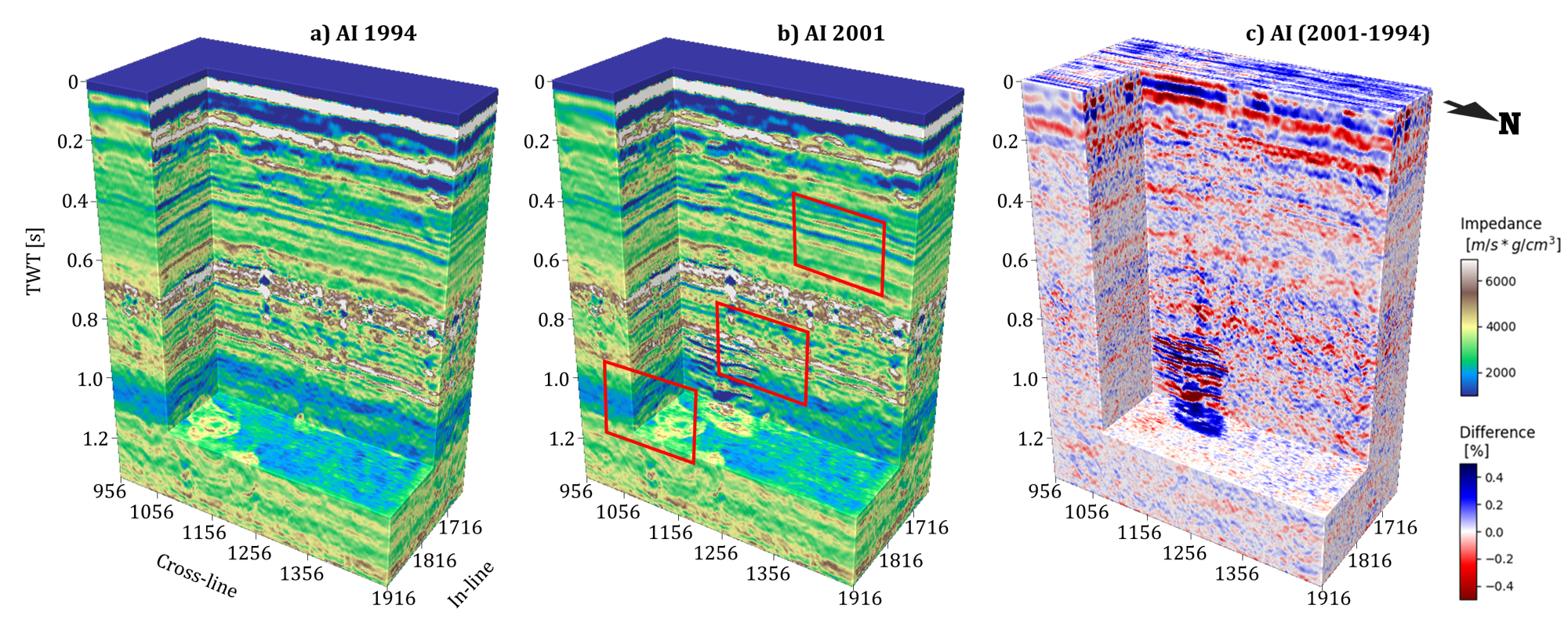}
    \caption{4D Tikhonov regularized inversion of the 1994 and 2001 Sleipner dataset vintages. a) Inverted baseline acoustic impedance model, b) inverted monitor acoustic impedance model, and c) the difference between a) and b). The red rectangles indicate the locations of the close-ups shown in Figure \ref{fig:windows}.}
    \label{fig:invl2}
\end{figure*}

\begin{figure*}[!h]
    \centering
    \includegraphics[width=\textwidth]{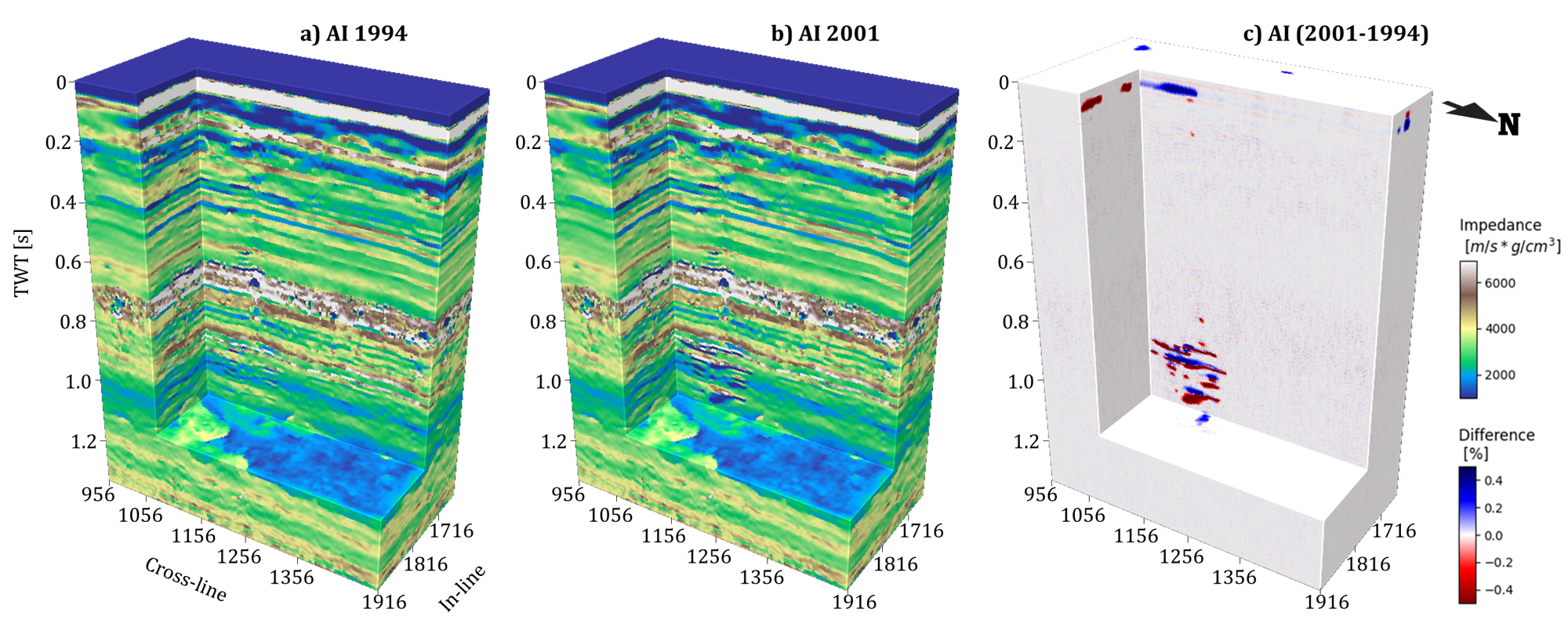}
    \caption{4D JIS of the 1994 and 2001 Sleipner dataset vintages. a) Inverted baseline acoustic impedance model, b) inverted monitor acoustic impedance model, and c) the difference between a) and b).}
    \label{fig:jis}
\end{figure*}

The baseline and monitor acoustic impedance models estimated from the 4D Tikhonov-regularized inversion, and 4D JIS methods are shown in Figures \ref{fig:invl2} and \ref{fig:jis}. For the Tikhonov regularized inversion, we use 100 iterations of the LSQR solver and set $\alpha=1.5$ and $\beta=0.1$ according to equation \ref{eqn:ip2}. In the case of 4D JIS, we use 2 outer iterations and 100 and 80 inner iterations for both Primal-Dual solvers in model parameter inversion and segmentation, respectively. The regularization parameters for 4D JIS are set to $\alpha = 0.2$, $\beta = 2$ and $\delta = 5$ (see Equation \ref{eqn:jis}).

\begin{figure*}[!h]
    \centering
    \includegraphics[width=5.8cm]{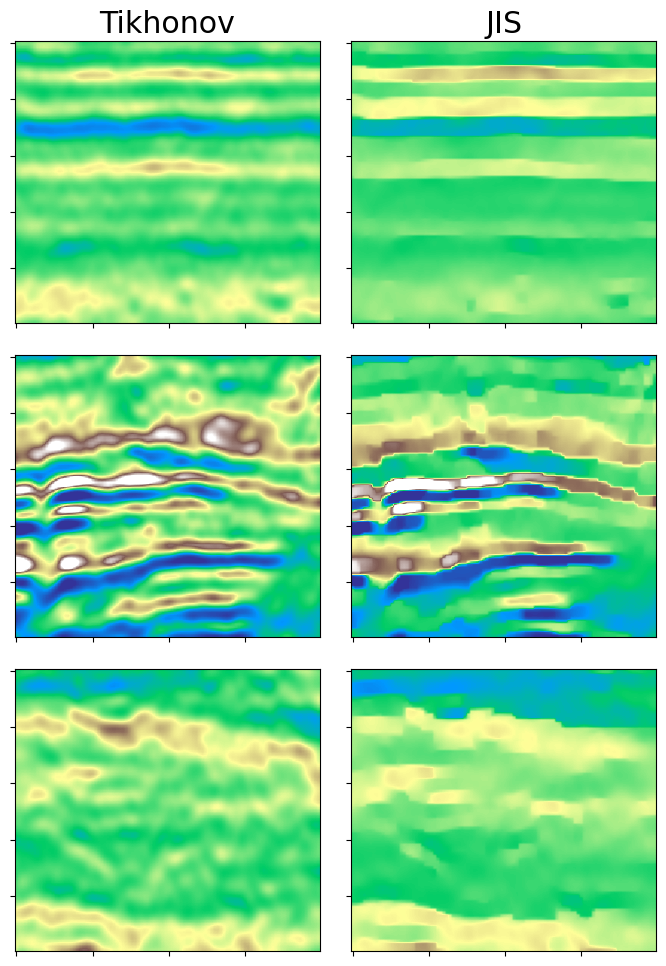}
    \caption{Close-ups that show a detailed view of the inversion results for the 4D Tikhonov regularized inversion (left column of images) and 4D JIS (right column of images).}
    \label{fig:windows}
\end{figure*}

We observe that the 4D JIS algorithm is able to obtain high-resolution acoustic impedance models with well-resolved layers (as shown in detail in figure \ref{fig:windows}) and considerably less non-repeatable noise than their Tikhonov-regularized counterparts. The superiority of 4D JIS over Tikhonov-regularized inversion is also clearly visible when comparing the 4D difference between the 1994 and 2001 seismic datasets (Figures \ref{fig:invl2}c and \ref{fig:jis}c): 4D JIS retrieves a difference volume with a markedly higher signal-to-noise ratio, highlighting the real time-lapse changes at the reservoir interval. Only a few small unrealistic time-lapse changes are retrieved by 4D JIS close to the seabed and in some parts of the overburden near the reservoir. We believe these changes to be non-geological and likely caused by amplitude mismatches in the input data, most likely a consequence of differences in acquisition and processing between the two datasets.

\begin{figure*}[!h]
    \centering
    \includegraphics[width=6cm]{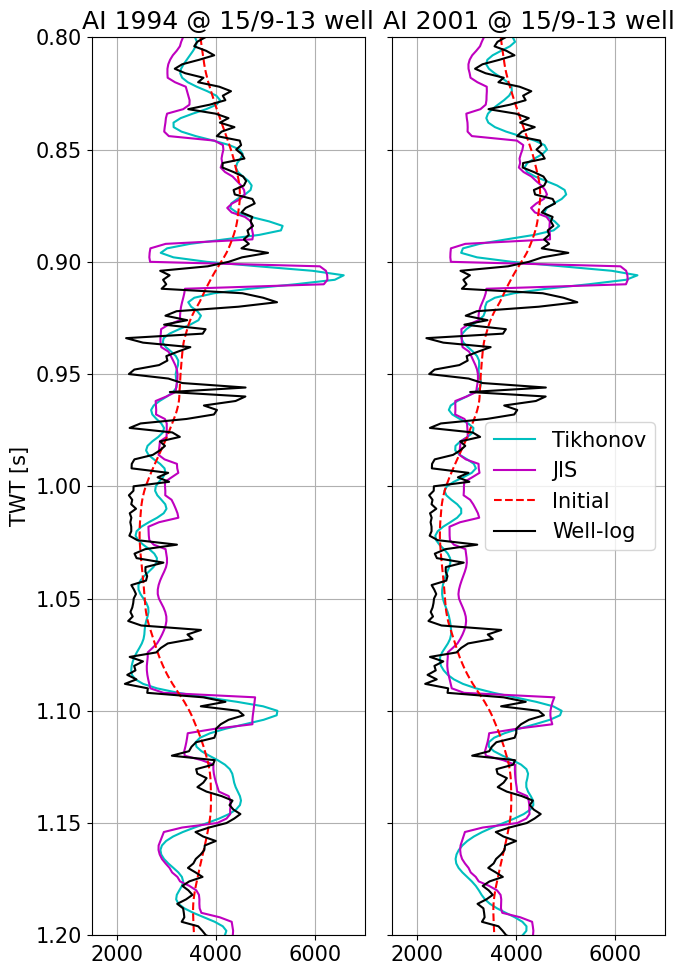}
    \caption{4D Tikhonov regularized inversion and 4D JIS results at the well location compared to the impedance well-log and the well-derivated initial velocity model.}
    \label{fig:well_comp}
\end{figure*}

Finally, the 4D inversion results are validated with well-data (Figure \ref{fig:well_comp}). These plots show that both 4D Tikhonov regularized inversion and our 4D JIS have a fair match with the impedance well-log at the reservoir interval ($0.9$ to $1.2$ ms). Whilst the results of 4D JIS are very close to those of Tikhonov regularized inversion, we can observe how some layers are better defined with sharper jumps at the known geological boundaries. 

\begin{figure*}[!h]
    \centering
    \includegraphics[width=7cm]{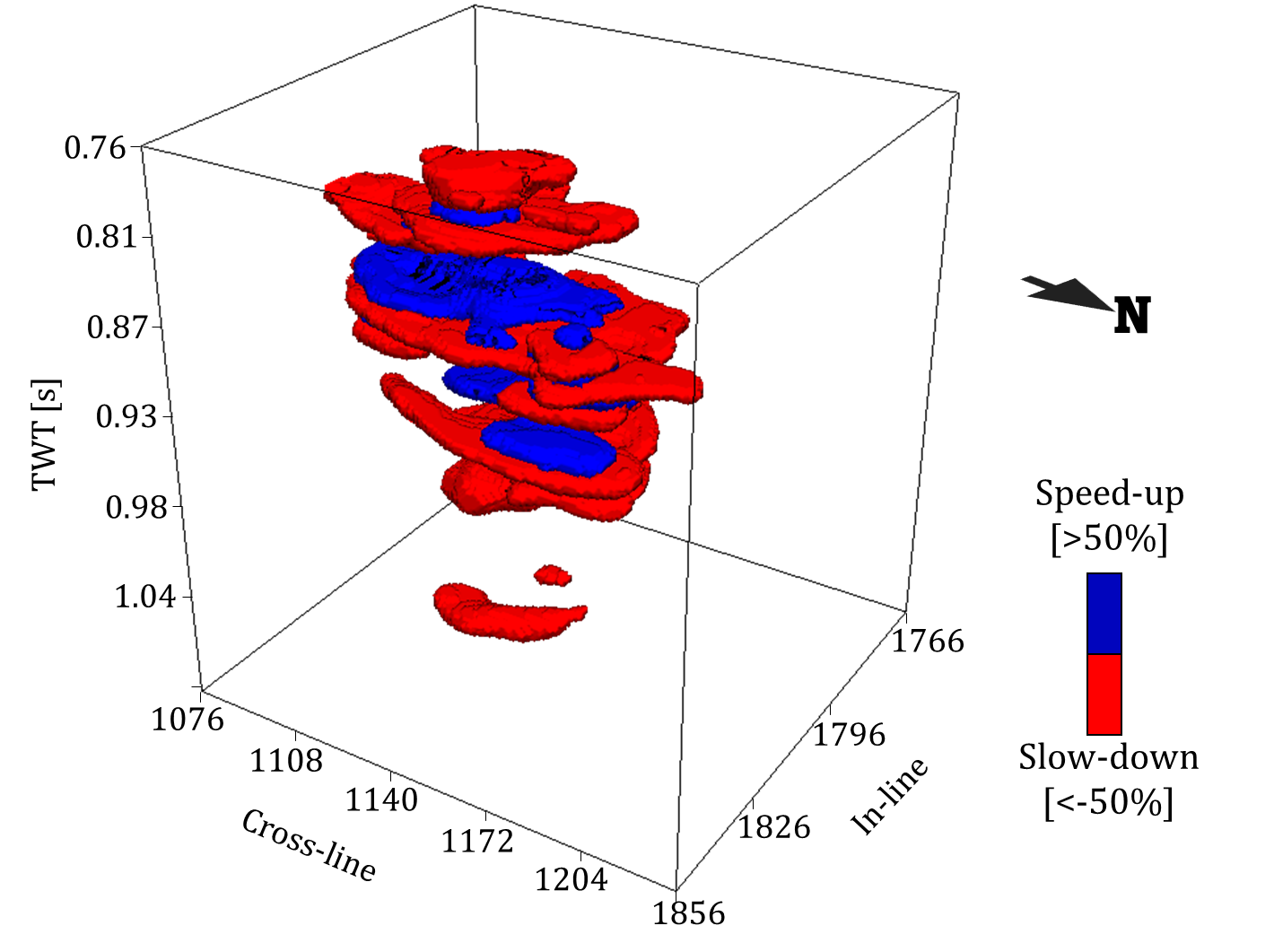}
    \caption{Volumetric model of the segmented time-lapse changes at Sleipner.}
    \label{fig:3D}
\end{figure*}

Additionally, JIS computes a segmented version of the baseline-monitor difference model (Figure \ref{fig:3D}). This volumetric model provides easy access to those cells that underwent an increase or a decrease in the acoustic impedance. The presence of both phenomena at the reservoir interval might suggest that the interbedded shale and sandstone rock layers are experiencing compaction and dilation, respectively, caused by the CO$_2$ injection that increases the pore pressure at the sandstone layers. However, it is important to consider that these values need to be corrected as part of the response might still be affected by amplitude tuning.

To further validate the quality of the segmented partitions, we calculate the overall volume of the geobodies corresponding to a $-50\%$ decrease in acoustic impedance and use it to estimate the mass of injected CO$_2$ at subsurface conditions. We assume porosity to be randomly distributed in the reservoir and set its mean value to $38\%$ as reported in the literature based on well-core, cuttings, and well-logs analyses \cite{Chadwick2004}. For the gas saturation, we use a constant value taken from the range of values estimated in \cite{Dupuy2017} based on rock physics analysis, namely $40\%$. As for the CO$_2$ density, we take a mean value at reservoir conditions of $720 kg/m^3$ \cite{Dupuy2017}. These values yield a CO$_2$ mass of $3.8$ Mt, corresponding to a slight underestimation given that the injected mass reported in 2001 was $4.2 Mt$. This underestimation can be caused by the fact that around $10\%$ of the injected CO$_2$ might get dissolved in the aquifer, and therefore it would be invisible to seismic methods due to the low contrast in acoustic impedance.

Lastly, regarding computer efficiency, Table \ref{tab:comptime} summarizes the time the 4D JIS algorithm takes to run both on a single CPU and on a GPU. Note that the first iteration of inversion is usually faster as it does not contain the additional segmentation term. From this result, we can conclude that the benefit of implementing such kind of scientific algorithms on GPUs is evident (no more or no less than what we are familiar with nowadays for deep learning applications).
\begin{table}[!h]
    \centering
    \def\arraystretch{1.5}
    \caption{Computation time of 4D JIS based on CPU and GPU. All experiments are run on an AMD EPYC
7713P 64-Core Processor equipped with a single NVIDIA TESLA A100}
    \begin{tabular}{ |c|c|c|c|c| } 
    \hline
     & & Inversion time $[min]$ &  Segmentation time $[min]$  \\
    \hline
    \multirow{2}{5.5em}{JIS on CPU} & Iteration 1 & 655.68 & 55.85 \\ 
     & Iteration 2 & 1747.82 & 56.10 \\
     \hline
     \multirow{2}{5.5em}{JIS on GPU} & Iteration 1 & 18.68 & 0.36 \\ 
     & Iteration 2 & 22.45 & 0.35 \\ 
    \hline
    \end{tabular}
    \label{tab:comptime}
\end{table}

\section{Conclusion}
In this work, we have applied the 4D joint segmentation-inversion (JIS) algorithm to the 1994 and 2001 seismic surveys of the 4D Sleipner seismic dataset. Our results show that 4D JIS outperforms the commonly used Tikhonov-regularized 4D inversion approach as it produces higher resolution acoustic impedance models by, at the same time, considerably reducing  the non-repeatable noise in the inverted models' difference. As a result, the subsurface changes due to CO$_2$ injection are clearly visible. Furthermore, JIS provides a segmented volume of the expected time-lapse changes that can ease 4D seismic interpretation and might be used to create 4D-driven geobodies for reservoir simulation models. By leveraging GPUs both in the application of the convolutional modelling operator as well a
s in the solution of millions of bisection problems within the segmentation step, we show that JIS can be easily scaled to large 3D data volumes, making it an appealing method for 4D seismic inversion in CO$_2$ monitoring projects. 

\section*{Acknowledgments}
The authors thank King Abdullah University of Science \& Technology (KAUST) for supporting this research  as well as Equinor and partners for releasing the 4D Sleipner dataset (available at \url{https://co2datashare.org/dataset/sleipner-2019-benchmark-model}).  For computer time, this research used the resources of the Supercomputing Laboratory at KAUST in Thuwal, Saudi Arabia.  The accompanying code to reproduce the results in this paper can be found at \url{https://github.com/DIG-Kaust/4DProximal}.'

\bibliographystyle{unsrt}  
\bibliography{references}

\end{document}